\documentclass[10pt]{iopart}
\usepackage{amsfonts,amssymb,mathrsfs}

\newcommand{\be}{\begin{equation}}
\newcommand{\ee}{\end{equation}}
\newcommand{\bea}{\begin{eqnarray}}
\newcommand{\eea}{\end{eqnarray}}
\newcommand{\nn}{\nonumber}
\begin{document}

\title{Modified gravity with $R$-matter couplings and (non-)geodesic motion}
\author{Thomas P. Sotiriou${}^1$ and Valerio Faraoni${}^2$}
\address{${}^1$ Center for Fundamental Physics,  University of Maryland, College Park, MD 20742-4111, USA\\
${}^2$Physics Department, Bishop's University, 2600 College St., Sherbrooke, Qu\`{e}bec, Canada 
J1M~1Z7}
\eads{\mailto{sotiriou@umd.edu}, \mailto{vfaraoni@ubishops.ca}}
\date{\today} 

\begin{abstract} 
We consider alternative theories of gravity with a direct 
coupling between matter and the Ricci scalar  We study the 
relation between these theories and 
ordinary scalar-tensor gravity, or scalar-tensor theories which 
include non-standard couplings between the scalar and matter. We then analyze the motion of matter in such 
theories, its implications for the Equivalence Principle, and 
the  recent claim that they can alleviate the dark matter 
problem in galaxies. 
\end{abstract} 

\section{Introduction}

An explanation for the nature of the unknown form of energy, 
usually dubbed dark energy, which appears to constitute 76\% of 
the energy budget of the universe \cite{Spergel:2006hy} and is 
responsible for its late time accelerated expansion is still 
pending. The dominant role of gravity in the dynamics of the 
late time universe and the unexpected characteristics of dark 
energy, namely the fact that it violates the Strong Energy 
Condition and may even be as exotic as phantom energy, have lead 
some to the idea that dark energy might be 
nothing but an indication of the inability of General Relativity 
(GR) to describe gravity consistently at cosmological scales 
\cite{firstmodified}.  
According to this approach, some other theory of gravity, 
which actually constitutes an extension of GR, might correctly 
describe the cosmological dynamics without dark energy and, at 
the same time, reduce to the usual GR phenomenology at scales 
where the latter has already been very successful.

Several theories have been put forth, often as toy models 
with the purpose of providing proofs of principle that such an 
explanation to the 
recent puzzles of cosmology can indeed be feasible. Most of them 
also seem to draw some motivation from high energy physics, as 
they are presented as candidates for a low energy 
effective action  of some more fundamental theory ({\em e.g.},  
String Theory). A 
typical example is that of the so-called $f(R)$  
gravity \cite{buchdahl}.  These theories, which are proposed as 
minimal extensions of GR, are described by the action 
\be \label{faction} 
S_f=\int d^4 x \sqrt{-g} \left[ \frac{1}{2\kappa} f(R) 
+ L_m(g_{ab},\psi) \right] \;,
\ee 
where $g$ is the determinant  of the 
metric $g_{ab}$, $\kappa=8\pi\, G$, $f(R)$ is a general 
function of the Ricci scalar $R$, $L_m$ is the matter 
action,  
and $\psi$ collectively denotes the matter fields (we 
adopt the notations and conventions of Ref.~\cite{Wald}). When 
$f(R)$ is taken to be linear, this action reduces to the usual 
Einstein-Hilbert action. Then, Einstein's equations can be 
derived with standard metric variation, but also with an 
independent variation of the metric and of the connection 
\cite{buchdahl, Wald} called Palatini variation, under the 
crucial assumption that the matter action does not depend on 
the, now independent, connection \cite{Sotiriou:2006sr}. 
Nevertheless, when $f(R)$ is not linear in $R$, the two 
variational principles lead to different theories: metric and 
Palatini $f(R)$ gravity \footnote{In the Palatini formalism, the 
connection $\Gamma^a_{\phantom{a}bc}$ is independent 
of the metric. Therefore, $R$ is replaced in the 
action~(\ref{faction}) by  ${\mathcal R} \equiv g^{ab} {\mathcal 
R}_{ab}$, where  ${\mathcal R}_{ab}$
   is the Ricci tensor constructed with this independent 
connection.} respectively (see \cite{buchdahl,ffv,Nojiri:2006ri, 
Capozziello:2007ec,Sotiriou:2007yd} for early works and 
reviews).  Additionally, one could decide to allow the 
independent connection to enter the matter action when the 
Palatini variation is used (which appears to be in accordance 
with the geometrical meaning of a connection to define the 
covariant derivative \cite{Sotiriou:2006sr}), in which case the 
outcome is yet another distinct version of $f(R)$ gravity: 
metric-affine $f(R)$ gravity \cite{Sotiriou:2006qn}.

Much attention has been paid to $f(R)$ theories of gravity 
lately, especially concerning the cosmological evolution but 
also the various viability constraints \cite{list} (with 
metric-affine $f(R)$ gravity being probably the exception). 
An extension of metric $f(R)$ gravity has been recently 
proposed in Ref. \cite{BBHL}, in which the Ricci scalar $R$  
acquires an explicit coupling to the matter Lagrangian. The 
action of the theory is 
\be \label{1} 
S=\int d^4x\, \sqrt{-g} 
\left\{ \frac{f_1(R)}{2} +\left[ 1+\lambda f_2(R) \right] L_m 
\right\} \;, 
\ee 
where $L_m$ is the matter Lagrangian, $f_{1,2} 
$ are ({\em a priori} arbitrary) functions of the Ricci 
curvature $R$, and the usual coefficient of $R$ in the 
Einstein-Hilbert action ($8\pi G$, where $G$ is Newton's 
constant) has been absorbed in $f_1(R)$.
 
This action, and variants of it, were considered for 
various purposes: in Ref.~\cite{BBHL}, (\ref{1}) was studied 
as an 
alternative to dark matter because it is claimed to give rise 
to  
phenomenology similar to MOND gravity \cite{MOND} on galactic 
scales; in \cite{Odintsov,Nojiri:2006ri}, a variation 
of~(\ref{1})  was 
studied as  an alternative to dark energy by setting $f_1(R)=R$ 
and keeping only the nonminimal coupling of matter to the Ricci 
curvature, in order to explain the current acceleration of the 
universe. Along the same line of thought, the idea of making 
the kinetic term of a (minimally 
coupled) scalar field dependent on the curvature, while keeping 
$f_1(R)=R$ was exploited in attempts to 
cure the  cosmological constant problem \cite{ShinjiRandall, 
DolgovKawasaki} (see also \cite{otherRkinetic, Sadjadi}). 
In light of its relation with a scalar-tensor theory 
explained below, the action~(\ref{1}) is reminiscent 
also of the one describing a nonminimally coupled 
electromagnetic field \cite{nonminimalem}.

Variation of the action (\ref{1}) with respect to the metric 
yields
\bea
\label{field}
& &F_1(R) R_{ab}-\frac{1}{2} f_1(R)  
g_{ab}-\nabla_a\nabla_b F_1(R)+g_{ab}\Box F_1(R)
\nn\\
&&\nonumber \\
&& = -2\lambda F_2(R) L_m R_{ab}+2\lambda 
(\nabla_a \nabla_b -g_{ab}\Box) L_m F_2(R)\nn\\
&&\nonumber \\
&&\qquad+[1+\lambda f_2(R)]  \, T_{ab} \;,
\eea
where $F_{1,2}(R)\equiv f'_{1,2}(R)$, the prime denotes 
differentiation with respect to the argument, and 
\be
T_{ab}\equiv -\frac{2}{\sqrt{-g}}\frac{\delta (\sqrt{-g} 
L_m)}{\delta g^{ab}}.
\ee
As a consequence of the explicit coupling of the matter 
Lagrangian density to the Ricci curvature, there is an 
energy exchange between matter and gravity beyond the usual one 
always 
present in curved spaces, which is reflected in the 
non-vanishing of the covariant divergence of the 
matter stress-energy tensor $T_{ab}$. The corrected conservation 
equation 
assumes the form \cite{BBHL, Odintsov, Nojiri:2006ri}
\be\label{conservation}
\nabla^b T_{ab}=\frac{\lambda F_2}{1+\lambda f_2}\left(  
L_m g_{ab}-T_{ab} \right) \nabla^b R \;.
\ee

The fact that the stress energy tensor is manifestly not 
divergence-free (in the representation used to write down the 
action~(\ref{1})) can be interpreted as a violation of the 
so-called metric postulates \cite{willbook}. This seems to imply 
that the theory exhibits violations of the Equivalence Principle 
(EP). This, as well as the fact that one could potentially tune 
the 
parameter $\lambda$  to reduce the effects of such 
violation below current experimental accuracy, has already been 
mentioned in Ref.~\cite{BBHL}. Nevertheless, to the best of our 
knowledge, no detailed study of this feature of the 
theories described by~(\ref{1}) has been carried out so far. 
Even though the metric postulates can be a good criterion for 
constructing theories which do not violate the EP, one can not 
safely conclude that any theory that does not satisfy the metric 
postulates in some particular representation indeed violates the 
EP. First of all, the metric postulates are 
representation-dependent  statements, {\em i.e.}, not invariant 
under field 
redefinitions \cite{TVS}. In general, there might be other 
representations of the theory in which the metric postulates are 
indeed satisfied \footnote{A characteristic example is 
scalar-tensor gravity: the metric postulates are satisfied in 
the 
Jordan frame but not in the Einstein frame. However, there are 
no violations of the Einstein EP, since the 
latter is a characteristic of the theory, not of its 
representations \cite{TVS} (see \cite{SF, think} for 
a discussion of the Einstein EP in 
nonminimally coupled scalar field theory, which is a particular 
realization of 
scalar-tensor gravity). It is just that in one frame this is 
made manifest.}. More importantly, even in theories that indeed exhibit violations of the EP, such as those considered here, the metric postulates or the non-conservation of energy do not themselves provide quantitative estimates of the deviations from the EP.

The purpose of this paper is exactly to shed some light on 
these issues. We begin by examining, in section~\ref{st}, the 
relation between the theories described by the action~(\ref{1}), 
ordinary scalar-tensor gravity, and scalar-tensor theories with 
an anomalous coupling between matter and the scalar field. 
It is shown that the former can not be cast into the form of any 
of the latter (as usually done in ``ordinary'' $f(R)$ gravity 
without the 
$R$-matter coupling). This  not only implies that the theory 
under consideration can not be cast away as a known theory, but 
also that the $R$-matter coupling can not be eliminated with 
simple field redefinitions. We proceed, in section 
\ref{Rmatter+ep}, to examine the effect of the $R$-matter 
coupling on the motion of particles and fields. It is shown that 
massless particle trajectories and field propagation are 
actually unaffected (at least 
at high frequencies). More remarkably, even the motion of 
massive objects (such as perfect fluids or test particles) seems 
to remain  unaffected. This 
implies that  detecting deviations from geodesic motion in the 
theories under 
investigation might be more difficult than expected. On the 
other hand, it also casts doubts on whether they can actually 
account for the phenomenology for which they were introduced in 
\cite{BBHL}.

\section{Is $R$-coupled modified gravity a dej\`a vu?}
\label{st}

It is customary in metric (and also Palatini) $f(R)$ 
gravity to introduce auxiliary scalar fields, in order to 
re-write the action as a scalar-tensor 
theory \cite{STequivalence}. Let us check what would happen if 
we 
followed the same procedure here: we begin by  
introducing two scalars $\Psi$ and $\phi$ and considering 
 the action 
\be \label{2a} S'=\int d^4x\, \sqrt{-g} 
\left\{ \frac{f_1( \phi)}{2}
 +\left[ 1+\lambda f_2( \phi) \right] L_m +\Psi(R-\phi)\right\} 
\;. \ee The variation of this action with respect to $\Psi$ 
yields $R=\phi$ and one recovers the action~(\ref{1}). Now, 
varying~(\ref{2a}) with respect to $\phi$ yields 
\be \label{psi} 
\Psi=\frac{1}{2} f'_1+\lambda f'_2 L_m \;. 
\ee 
Assuming that at  least one of $f_{1,2}$ is non-linear in $R$ 
($f''_1\neq 0$ or 
$f''_2\neq 0$, or both) \footnote{The particular case in which 
both 
functions are linear in $R$, $f''_1=f''_2=0$, is burdened with  
serious viability issues  \cite{thomasBBHL} and will not be 
studied further.} and 
replacing eq. (\ref{psi}) back in eq. (\ref{2a}) one gets (see 
also \cite{Odintsov, Nojiri:2006ri}) 
\bea \label{2} S'& 
=& \int d^4x\, 
\sqrt{-g} \Bigg\{ \frac{f_1( \phi)}{2} + \left[ 1+\lambda f_2( 
\phi) \right] L_m\nn\\ & &\quad\qquad\qquad+ \left( \frac{1}{2}f'_1+\lambda f'_2 
L_m\right)\left( R-\phi \right)\Bigg\} \;. \eea 
This action is 
obviously dynamically equivalent to the action~(\ref{1}). In the 
special case $\lambda=0$, in which~(\ref{1}) reduces to 
the action~(\ref{faction}), and~(\ref{2}) reduces to 
\bea 
S'=\int 
d^4x\, \sqrt{-g} \Bigg\{ \frac{f_1( \phi)}{2} +L_m+ 
\frac{1}{2}f'_1\left( R-\phi \right)\Bigg\} \;, 
\eea 
or, with 
the simple field redefinition $\Phi \equiv f'_1(\phi)$ and the 
introduction of the potential $V(\Phi)=(\phi f'_1-f_1)/2$, it 
can be rewritten as a Brans-Dicke theory with a potential and  
vanishing Brans-Dicke parameter $\omega_0$ \cite{STequivalence}: 
\be 
S'=\int d^4x\, \sqrt{-g} \Big\{ \frac{\Phi R}{2} -V(\Phi) 
+L_m \Big\} \;. 
\ee 
However, when $\lambda\neq 0$, one can 
see directly  from the action~(\ref{2}) that the coupling 
between 
matter and $R$ does not disappear. Even if we perform the field 
redefinition mentioned earlier and introduce the functions 
$V(\Phi)$, $ U(\Phi) \equiv 1+\lambda (f_2-\phi f'_2)$, and 
$X(\Phi) \equiv f'_2$, at best we can write \be \label{final} 
S'=\int 
d^4x\, \sqrt{-g} \Big\{ \frac{\Phi R}{2} -V(\Phi) + U(\Phi) 
L_m+\lambda X(\Phi) R L_m \Big\} \;. \ee
  This is clearly not an 
ordinary scalar-tensor theory: first, there exists an unusual 
coupling between the scalar and the matter in the third term. This 
is reminiscent of extensions of scalar-tensor gravity which 
include similar couplings, such as the theories considered by 
Damour and Polyakov \cite{Damour:1994zq}. However, the presence 
of the $R$-matter (-scalar)  coupling in the last term 
distinguishes the action (\ref{final})  even from these 
generalized scalar-tensor theories.  Alternative ways to 
introduce a single scalar are also unable to alleviate 
the 
$R$-matter coupling (see the Appendix). 

Also, one should not be mislead to think that the $R$-matter coupling can be eliminated, judging from action (\ref{2a}). Even though in eq.~(\ref{2a}) there is indeed no explicit coupling between $R$ and the matter, there is a coupling between $\Psi$ and $R$. $\Psi$ in turn is just an auxiliary scalar which is algebraically related to the matter fields through eq.~(\ref{psi}). Consequently, the $R$-matter coupling is just made implicit through the introduction of an auxiliary (non-dynamical) field.
Note the difference with $\phi$ (or $\Phi$) which does carry dynamics (it is algebraically related to $R$ but not to the matter).

Therefore, we conclude the following:
\begin{itemize}
\item  Introducing scalar fields helps in avoiding 
the presence of non-linear functions of $R$ in the action (as in ordinary $f(R)$ gravity).
\item However, the theories under scrutiny cannot be written into the form of a scalar-tensor theory with a minimal coupling to matter.\footnote{By scalar-tensor theory we mean a theory with a single scalar field mediating gravity. For theories with more than one scalars we use the term multi-scalr-tensor theory.}
{\em I.e.}~the $R$-matter coupling cannot be eliminated, or replaced by some 
unusual couplings between matter and the scalar field (it is, therefore, not an artifact of some peculiar representation).
\item One could ``hide'' the $R$-matter coupling by considering the theory as a multi-scalar tensor theory (starting from action (\ref{2a}) and possibly considering also field redefinitions for the two scalars). However, in this case special attention should be paid to the fact that one of these scalar does not actually carry dynamics. This, for instance, distinguishes this theory from the multi-scalar-tensor theories already considered in the literature \cite{Damour:1992we}.
\end{itemize}
Consequently, the theories described by the action~(\ref{1}) can 
not be cast into the form of a scalar- or multi-scalar-tensor gravity
 which has already been extensively studied in the literature and its phenomenology is well known: The presence of the $R$-matter coupling  which is exactly the subject of interest here, is what distinguishes these theories and deserves further attention.

\section{$R$-coupling, geodesics, and EP violations}
\label{Rmatter+ep}

As already mentioned, the explicit coupling to the Ricci  
curvature described by the action~(\ref{1}) can potentially lead to
non-geodesic motion and violations of the EP,  judging from the 
fact that the stress-energy tensor is not 
divergence-free, as expressed by eq.~(\ref{conservation}). 
Although this is pointed out in Ref.~\cite{BBHL} with the 
caveat that such violations could be controlled by the value of 
the  parameter $\lambda$, this issue has not yet been addressed 
in detail. In this section we  perform a more thorough analysis. 
First, we  consider massless particles and high 
frequency fields, and then we 
move on to perfect fluids.  

\subsection{Massless fields}

In~\cite{BBHL}, it is stated that the explicit coupling of the 
matter Lagrangian to the Ricci curvature does not alter the 
{\em null} 
geodesic equation: however, this statement is not proved, and 
the 
reader might be left with the impression that it is 
motivated by the resemblance of the action~(\ref{1}) with the 
one 
of  Einstein 
frame scalar-tensor theories, which we have shown  to be 
fallacious. As a matter of fact, the authors of~\cite{BBHL} are 
more interested in the correction to the 
worldlines of {\em massive} particles and the corresponding 
MOND-like 
phenomenology; however, this aspect of the theory should not 
be left unchecked.

\subsubsection{Null dust:}

The equation for null geodesics can be derived in a  
straightforward manner from the conservation equation of a null 
dust fluid (see {\em e.g.}, Ref.~\cite{nulldust}). Therefore, 
if the $R$-coupling were to induce any 
corrections to the null geodesic equation, these would show up 
in this derivation. 

The stress-energy tensor of  a null dust is
\be
T_{ab}=\rho \, u_a \, u_b \;,\;\;\;\;\;\;\; u_cu^c=0 \;,
\ee
where $\rho$ is the fluid energy density and $u^c$ is the 
four-velocity of massless fluid particles. Since the perfect fluid is an 
``averaged'' and not an exact description   for matter, it is more common 
in this case to work directly with the  stress-energy tensor and avoid any 
reference to a Lagrangian. Unfortunately,  in our case this is not 
possible since $L_m$ enters explicitly in  the conservation equations 
(\ref{conservation}). Even though more than  one choices for the 
Lagrangian have been used in the literature  
 \cite{SchutzBrown,HawkingEllis}, the most natural (and general) choice 
seems to be  simply the pressure $L_m=P$  
\cite{SchutzBrown}.\footnote{Note that the choice of the Lagrangian is 
particularly meaningful here: in GR these different choices would anyway lead to the same field equations, unlike here where the actual expression for $L_m$ enters the field equation of the theory explicitly. Therefore, all results strongly depend on this choice.}  The corrected conservation 
equation~(\ref{conservation})  then becomes
\be \label{nulldust}
\rho u^b \nabla_b  u_a 
+ \rho u_a \nabla^b  u_b 
+ u_a u_b \nabla^b  \rho  
=-\frac{\lambda 
F_2}{1+\lambda f_2} \, \rho u_au^b \nabla_b R \;.
\ee
Now this equation can be written in the form
\be
\label{nonaff}
u^b \nabla_b  u_a =\Theta u_a \;,
\ee
where
\be
\Theta=-\nabla^b  u_b +\frac{1}{\rho} \, u_b \nabla^b  \rho  
-\frac{\lambda 
F_2}{1+\lambda f_2} \, u^b \nabla_b R \;.
\ee
Eq.~(\ref{nonaff})  states that the  four-velocity is 
parallel-transported along the path: this is  the definition of 
a geodesic curve. The fact that this equation differs from the 
more familiar form $u^c\nabla_c u^a=0$ is simply because the 
latter is affinely parametrized. Therefore,  
the explicit $R$-coupling does not change the equation of null 
geodesics.

\subsubsection{Massless scalar field:}

Let us consider now a massless scalar field $\phi$ described by 
the Lagrangian density and energy-momentum tensor 
\cite{Chernikovetal}
\begin{eqnarray}
&& L_m =-\frac{1}{2} \, \nabla_c\phi\, \nabla^c\phi \;, 
\\
&&\nonumber \\
&& T_{ab}=\nabla_a\phi\, \nabla_b\phi -\frac{1}{2} \, g_{ab} 
\nabla_c\phi\, \nabla^c\phi  \;.
\end{eqnarray}
With a  correction proportional to $g_{ab} 
L_m-T_{ab}=-\nabla_a\phi \nabla_b \phi$, the (non-)conservation 
equation~(\ref{conservation}) is 
\begin{eqnarray}
&& \left( \nabla^b\nabla_a \phi \right) \nabla_b \phi +\left( 
\nabla_a \phi \right) \Box \phi -\frac{1}{2} \left( \nabla_a 
\nabla^c\phi \right) \left( \nabla_c \phi \right) -\frac{1}{2} 
\, \left( \nabla^c\phi \right) \nabla_a\nabla_c\phi \nonumber \\
&&\nonumber \\
&& =  -\, 
\frac{\lambda F_2}{1+\lambda f_2}\left( \nabla_a\phi \right) 
\left( \nabla_b \phi \right)\nabla^b R \;.
\end{eqnarray}
By projecting onto $\nabla^a\phi$, one obtains the corrected 
Klein-Gordon equation
\be\label{correctedKG}
\Box \phi + \,  \frac{\lambda F_2}{1+\lambda f_2}\left( 
\nabla_b\phi 
\right) \nabla^b R =0 \;.
\ee
By taking the high frequency limit 
\be
\phi \left( x^a \right)=\phi_0\left( x^a\right) \, 
\mbox{e}^{iS(x^c)} \;,
\ee
with the phase $S$ a rapidly varying function of $x^c$ and the 
amplitude $\phi_0\left( x^a\right)$ slowly varying, and 
neglecting the gradients and second derivatives of $\phi_0$, 
eq.~(\ref{correctedKG}) implies
\be
-S_a S^a +i\nabla^c S_c=
-\,  \frac{\lambda F_2}{1+\lambda f_2} \, i S_b \nabla^b R \;,
\ee
where $S_c\equiv \nabla_c S$ is the phase gradient and the 
tangent to the  worldline of the scalar particle in the 
geometric optics approximation. This 
equation yields 
\be \label{nullsf}
S_a S^a=0 
\ee
({\em i.e.}, the spacetime trajectory is null), and
\be
\nabla^c S_c=-\,  \frac{\lambda F_2}{1+\lambda f_2} \, S_b 
\nabla^b 
R \;,
\ee
which expresses the fact that the ``scalar photon'' is not 
transversal unless $\lambda =0$ or $S^b$ is orthogonal to the 
gradient of $R$. By further applying the covariant derivative 
operator to eq.~(\ref{nullsf}) and using the fact that $\left[ 
\nabla_a, \nabla_b \right]S=0$, one obtains 
\be
S^a \nabla_a S^b=0 \;.
\ee
Therefore, the worldlines of scalar particles  in the high 
frequency approximation are {\em null geodesics}.
This conclusion would not change if one were to consider 
instead a massless, non-minimally coupled, scalar field 
described by the Lagrangian density
\be
L_m =-\frac{1}{2} \, \nabla_c\phi\, \nabla^c\phi 
-\frac{\xi}{2} \,  R\phi^2 \;,
\ee
where $\xi$ is a dimensionless coupling constant. In fact, the 
corresponding Klein-Gordon equation
\be 
\Box \phi   -\xi  R \phi = 
-\,  \frac{\lambda F_2}{1+\lambda f_2}\left( \nabla_b\phi 
\right) \nabla^b R 
\ee
contains the extra term $-\xi R \phi$ that can be interpreted as 
a tidal effect 
on the scalar field which is important only for long wavelengths 
and disappears in the high frequency limit, yielding again the 
null 
geodesic equation.

\subsubsection{Maxwell field:}

Let us now consider the Maxwell field with Lagrangian density 
and energy-momentum tensor
\begin{eqnarray}
&& L_m =-\frac{1}{16\pi} \, F_{cd}F^{cd}
\;, \\
&&\nonumber \\
&& T_{ab}=\frac{1}{4\pi} \,  \left( F_{ac}{F_b}^c -\frac{1}{4} 
\, g_{ab} F_{cd}F^{cd} \right)  \;.
\end{eqnarray}
The corrected conservation equation~(\ref{conservation}) yields
\be
\nabla^b 
\left( F_{ac}{F_b}^c -\frac{1}{4} 
\, g_{ab} F_{cd}F^{cd} \right) =
-\,  \frac{\lambda F_2}{1+\lambda f_2} \, F_{ac}{F_b}^c \nabla^b 
R \;.
\ee
By taking now the high frequency limit
\be
A_b=C_b\, \mbox{e}^{iS}
\ee
with the phase $S(x^c)$ a rapidly varying function and $C_b $ a 
slowly varying vector amplitude, and neglecting the derivatives 
of the latter, one obtains (using 
$F_{ab}=\partial_aA_b-\partial_b A_a$),
\bea
2\alpha_{ab}S^b -S_a \left[ S^2 C^2 -\left( S_c C^c \right)^2 
\right]&=&0 \;, \label{this1}\\
\nabla^b\alpha_{ab}-\frac{1}{2} \nabla_a \left[ S^2 C^2 
-\left( S_cC^c 
\right)^2 \right] &=&- \frac{\lambda F_2}{1+\lambda f_2} \, 
\alpha_{ab}\nabla^b R \;, \label{this2}
\eea
where $S_c\equiv \nabla_c S , C^2 \equiv C_c C^c, S^2 \equiv S_c 
S^c$, and
\be
\alpha_{ab}= C^2 S_aS_b -\left( C^c S_c \right) \left( S_a C_b 
+S_b C_a \right) +S^2 C_a C_b  \;. \label{alpha}
\ee 
Eqs.~(\ref{this1}) and (\ref{alpha})  yield $ \left[ C^2S^2-\left( C_c S^c\right)^2 
\right] S_a=0$ and 
\be \label{A}
C^2S^2=\left( C_c S^c\right)^2
\ee
and eqs.~(\ref{this1}) and (\ref{this2}) simplify to
\begin{eqnarray}
&& 2\alpha_{ab}S^b =0 \;, \\
&& \nabla^b \left[ \left( 1+\lambda f_2 \right) \alpha_{ab}\right]=0 \;.
\label{thiiis}
\end{eqnarray}
The only difference with respect to the standard Maxwell equations in 
curved space is the term in $\lambda$ in eq.~(\ref{thiiis}). It 
is straightforward to see that, in this equation, 
\be
\left| \nabla^b \alpha_{ab} \right|\approx \left| 
\frac{\alpha_{ab}}{\lambda_{em}}\right| >> \left| \alpha_{ab} \nabla^b 
\left[ 1+\lambda f_2(R) \right] \right| \approx \left| 
\frac{\alpha_{ab}\lambda f_2}{L}\right| \;,
\ee
where $\lambda_{em} $ is the wavelength of electromagnetic waves and $L$ 
is the radius of curvature os spacetime. In the high frequency 
limit $\frac{\lambda_{em}}{L}<<1$, the corrections to standard optics 
coming from the term $\lambda f_2(R)$ in the Lagrangian disappear and 
photons follow null geodesics and are transversal. In other words, the 
non-minimal coupling corrections to the Maxwell equations can only 
affect long wavelenghts, comparable to the radius of curvature of 
spacetime. In the high frequency limit, photons are transverse and 
propagate along null geodesics.

\subsection{Perfect fluid with constant equation of 
state}

Let us now turn our attention to massive matter fields and, for 
simplicity, let us consider a perfect fluid  composed of 
non-relativistic or relativistic  
particles with constant  barotropic equation of state 
$P=w\rho$, where $\rho $ and $P$ 
are the energy density and pressure, respectively, described by 
the stress-energy tensor 
\be
T_{ab}=\left( w+1\right) \rho u_a u_b +w \rho g_{ab} 
\ee
and by the Lagrangian $L_m=P=w\rho$. The corrected 
conservation equation~(\ref{conservation}) yields
\bea
\label{pfcons}
&&\left( w+1\right) u_au^b \nabla_b \rho +\left( w+1\right)  
\rho  u^b \nabla_b u_a \nonumber\\
&&\nonumber \\
&& \quad + \left( 
w+1\right) u_a \nabla^b u_b+w\nabla_a 
\rho= -\frac{\lambda \left( w+1\right) F_2}{1+\lambda f_2}\, 
\rho u_au_b \nabla^b R \;.
\eea
Projecting onto the direction of the fluid four-velocity 
$u^a$, one obtains
\be
\frac{ D\rho}{D\tau}+\left(w+1 \right)\rho \nabla^b u_b 
=-\, \frac{\lambda \left( w+1 \right)F_2}{1+\lambda f_2} \, \rho 
\, 
\frac{DR}{D\tau} \;,
\ee
where $ \tau $ is the proper time along the timelike fluid 
curves 
and $D/D\tau \equiv u^c\nabla_c $. The correction to the 
conservation equation is best seen in the weak-field limit, in 
which $\tau \sim t$, $ \frac{D\rho}{D\tau} \simeq 
\frac{d\rho}{dt} =\frac{\partial \rho}{\partial t} +\vec{v}\cdot 
\vec{\nabla}\rho$, and 
\be\label{weakfield}
\frac{\partial \rho}{\partial t} +  \vec{\nabla}\cdot\left( \rho
\vec{v} \right) = -\, \frac{\lambda \left( w+1 
\right)F_2}{1+\lambda f_2} \, 
\rho \left(\frac{\partial R}{\partial t}+\vec{v}\cdot 
\vec{\nabla}R\right) \;.
\ee
Clearly, the fluid energy is not conserved and the right hand 
side of eq.~(\ref{weakfield}) acts as a source term describing 
the energy injected into the  fluid per unit time and per unit 
volume. This correction disappears if $\lambda\rightarrow 0$, in 
vacuo, quantum vacuo ($w=-1$), or when $DR/D\tau=0$.

We now project eq.~(\ref{pfcons}) onto the direction normal to 
the  four-velocity by the use of the projection operator 
${h^a}_c$ defined by  
$h_{ab} \equiv g_{ab}+u_{a}u_{b}$ (recall that, in 
our  signature, $u^c u_{c}=-1$). This gives, after easy 
manipulations,
\be
\frac{D u^a}{D\tau}\equiv \frac{du^a}{d\tau}+ 
\Gamma^{a}_{bc} u^b u^c =f^a \;,
\ee
where
\be
\label{extraforce}
f^{a}=\frac{1}{\rho(1+w)}\left[\frac{\lambda F_2}{1+\lambda 
f_2}(L_m-P)\nabla_c R+\nabla_c P\right] h^{ac} \;.
\ee
The last term in square brackets,  proportional to the pressure 
gradient, is the usual term that appears in GR  
and encapsulates the force exerted on a fluid element due to the 
fluid pressure (it is not to be attributed to the 
coupling  between matter and $R$, nor does it signal any new 
effect). As mentioned earlier, a natural choice for the 
Lagrangian of a  perfect fluid is $L_m=P$ \cite{SchutzBrown}. 
Substitution into eq.~(\ref{extraforce}) yields
\be
f^{a}=\frac{1}{\rho(1+w)}(\nabla_c P) h^{ac},
\ee
{\em i.e.}, remarkably, the only force is due to the pressure 
gradient already present in GR. For instance, in the case of 
dust with $P=0$, it is $f^c=0$. Therefore, we conclude that 
there is no extra force which can be attributed to the coupling 
between the matter and $R$ even for the case of a perfect fluid. 
The fact that energy is indeed not conserved for this fluid does 
not contradict this result. On the contrary, one can  
readily see from the  right hand side of eq.~(\ref{pfcons}) that 
the flow of energy only occurs along the direction  of $u^c$, 
{\em i.e.}, aligned with the fluid worldlines. Only the time  
component of the force is non-zero, while its spatial components 
in the frame comoving with the fluid always vanish. This kind of 
force can have no effect on the motion because, due to the 
normalization $u^c u_c=-1$, the only meaningful component of the 
four-acceleration $a^c$ and of the four-force  $f^c\propto 
a^c$ is the one perpendicular to the 
four-velocity ($a_c u^c=0$), {\em i.e.}, the spatial one. For a 
timelike four-velocity $u^c$, a four-acceleration parallel to 
$u^c$ is necessarily zero, which is what we recover here. In a   
Newtonian analogy, the correction to the equations of motion 
would correspond to the introduction of a spatially homogeneous, 
but time-dependent, potential energy.

The fact that no extra force appears for a perfect fluid 
was missed in Ref.~\cite{BBHL}. We will discuss the implications  
of this fact, to some extent, in section \ref{violation}. For 
the moment, let us stress the following: In \cite{BBHL} the 
authors use the opposite signature than the one used here, but  
also in Refs.~\cite{SchutzBrown} where the result $L_m=P$ is 
derived. The use of a different signature  is also the reason 
for the sign difference in the parenthesis  and in front of $P$ 
between eq.~(\ref{extraforce}) and the corresponding equation of
Ref.~\cite{BBHL}. Both equations are consistent given the 
signature adopted. However, when replacing $L_m$ in  this 
equation one has to take the signature into account as well: 
since  the gravitational action changes sign after a signature 
change, the sign of the matter action has to be changed for  
consistency, {\em i.e.}, $L_m=-P$ with the signature convention 
of \cite{BBHL}. Note also that in a follow-up publication 
by Bertolami and P\'{a}ramos studying the effects of the 
coupling discussed here on stellar configurations \cite{BBHL2}, 
it is assumed that $L_m=P$, even though the same signature as in 
Ref.~\cite{BBHL} is used, which is the opposite of the one used 
here and in Refs.~\cite{SchutzBrown}. This certainly affects the 
results derived and, therefore, one has to be cautious about the 
conclusions stated there until the effect of this inconsistency 
in the  analysis is assessed.

It is worth noting that the above conclusions apply to a 
massive scalar as well. This can be 
understood through the fact that a massive scalar admits a 
perfect fluid representation. Alternatively, it can be seen 
directly by the fact that the right hand side of eq. 
(\ref{conservation}) is proportional to $L_m 
g_{ab}-T_{ab}$. Given the definition of the stress 
energy tensor, any part of the Lagrangian $L_m$ which does not 
explicitly depend on the metric, such as the potential of a 
scalar field, does not contribute to this term. Massless or 
massive, a  scalar field will always lead to an energy flow 
along its motion 
only, resulting into a vanishing extra force.

\subsection{Conformal frames and energy conservation}

We have already discussed the fact that the $R$-matter coupling 
can not be eliminated by a conformal transformation of the 
metric 
for a generic matter Lagrangian. This is  evident in the 
fact that $R$ couples in the same way to all the terms that a 
matter Lagrangian might be split into (kinetic, 
potential, {\em etc.}). However, one could think of employing 
the 
tool of conformal  transformations  in order to see 
how particular matter 
fields are affected by this coupling. More specifically, 
consider a matter field whose Lagrangian does indeed transform 
as $L_m \rightarrow A L_m$ under the transformation 
$g_{ab}\rightarrow \Omega^2 g_{ab}$, where $A$ is some 
power of the scale factor $\Omega$. Then it does seem 
possible, under certain conditions, to find a conformal frame in 
which energy is conserved for this field (and, consequently, 
particles associated with this field  follow geodesics of this 
metric). Recall also that geodesic motion is a characteristic of 
the theory, not of its representations, so changing conformal 
frames in an effort to find one that makes it manifest is 
perfectly legitimate.\footnote{We are referring 
to whether a particle follows geodesics of some metric. The 
Einstein EP requires that 
the particles follow geodesics of some 
metric, not necessarily the one the action is written with. Of 
course, if particles corresponding to different fields follow 
geodesics of different metrics, then this signals a violation of 
the universality of free fall. Such a violation however, can 
only be  detected in E\"{o}tvos-type experiments by comparing 
the trajectories of different particles  and not by detecting 
deviation of a single species from 
geodesic motion (see section \ref{violation}).} Indeed, we can 
re-derive some of the results of the previous section with the 
use of conformal frames. This, besides being a way of verifying  
their validity, is also a nice exercise for realizing the 
ambiguities of the metric postulates, already pointed out in 
\cite{TVS}.

We begin by recalling that, if under the conformal 
transformation
\be
g_{ab}\rightarrow \tilde{g}_{ab}=\Omega^2 g_{ab}
\ee
the stress energy tensor $T_{ab}$ transforms as 
\be
\label{settrans}
T^{ab}\rightarrow \tilde{T}^{ab}=\Omega^s T^{ab} \;,
\ee
where $s$ is an appropriate conformal weight, then one can 
 easily verify that the following equation relates the 
covariant divergencies of $\tilde{T}_{ab}$ and $T_{ab}$ 
\cite{Wald}:
\be
 \tilde{\nabla}_a \tilde{T}^{ab}=\Omega^s \nabla_a
T^{ab}+(s+6)\Omega^{s-1} 
T^{ab}\nabla_{a}\Omega-\Omega^{s-1} T \nabla^b \Omega \;,
\ee
 where $\tilde{\nabla}_c$ is the covariant derivative 
 associated with $\tilde{g}_{ab}$ and 
$ T=g_{ab}T^{ab}$. Now, by imposing the condition that 
$\tilde{\nabla}_a \tilde{T}^{ab}=0$, {\em i.e.}, that 
energy is conserved in the new frame, we get
\be
 \nabla_a T^{ab}= \frac{1}{\Omega}\left[ T g^{ab} 
-(s+6) T^{ab}\right] \nabla_b\Omega \;. 
\ee 
We rewrite here  eq.~(\ref{conservation}) for direct comparison 
\be\label{conservation2} 
\nabla_b T^{ab}=\frac{\lambda 
F_2}{1+\lambda f_2}\left( L_m g^{ab}-T^{ab} \right) \nabla_b R 
\;. 
\ee 
One can then easily notice that these two equations 
become identical if and only if 
\be \label{cond} 
s=-4 \;, \qquad T=2  L_m \;, \qquad \Omega^2=1+\lambda f_2(R) 
\;. 
\ee 
The 
first two 
conditions give us characteristics of the matter fields and the 
last just pinpoints the appropriate choice of the conformal 
factor. The only case that we know of that  satisfies the 
first two conditions (considering also the transformation 
rule~(\ref{settrans})) is a perfect fluid with stiff equation 
of state $P=\rho$, a particular realization of which is a  
massless scalar field (which admits a perfect fluid 
representation). Therefore, 
if matter is only composed of such fluids or scalars, then a 
conformal frame can be found in which energy is conserved and 
the corresponding conformal factor is the one in 
eq.~(\ref{cond}).

We stress once more that the fact that this procedure can not be 
carried out for all matter fields directly implies that energy 
conservation is 
not a generic characteristic of the theory. Also, the fact that 
massless scalar particles follow geodesics on the one hand, or 
the  fact that we can not show geodesic motion for all fields on 
the 
other hand,  should not mislead us to think that we can make any 
statement (positive or negative, respectively) about the EP. In 
fact, we have 
already shown above that  also null dust, high-frequency 
photons, and perfect 
fluid volume elements 
follow geodesics, even though the use of conformal frames could 
not provide these results.

Since this is actually the lesson to be learned from this  
exercise, let us comment on it: In the  
perfect fluid (including null dust) case, we have already seen 
that 
energy is not conserved. Geodesic motion comes as a 
consequence of the fact that the energy flow is always timelike 
and parallel to the fluid worldlines instead of being 
purely spatial.  
For the case of the electromagnetic field, on the 
other hand, we already know that the matter action is 
conformally invariant, so one can not expect 
that much information can be obtained through the use of 
conformal frames. The bottom line is that reaching  
conclusions about geodesic motion and EP  
violations from energy conservation (the divergence of 
the  stress energy tensor in some conformal frame) is not always 
an  easy and unambiguous task as one may think when considering 
the metric postulates. This fact further supports the 
claims  made in \cite{TVS}.

\subsection{Equivalence principle violations and dark matter}
\label{violation}

\subsubsection{Possible violations of the Equivalence 
Principle:}

As already mentioned, the unusual coupling between $R$ and 
matter in the theories under investigation and the fact that 
energy is not conserved signal a violation of the EP. On 
the other hand, in order to get concrete evidence of that, but 
also in order to derive quantitative results, this issue should 
be examined further, since there are intricacies in using the 
metric postulates for judging  whether, and how, a theory will 
show such violations. First of all, let us be more concrete and 
separate the Strong Equivalence Principle (SEP), the 
Einstein Equivalence Principle (EEP), and the Weak Equivalence 
Principle (WEP) (see \cite{willbook} for definitions of these 
concepts).

$R$ couples in the same manner to all possible matter Lagrangians into which one could decide to split $L_m$ (one for each matter field. Therefore, there is no reason to believe that test particles 
(which by definition do not contribute to the gravitational 
field) of different compositions will follow different 
trajectories. This implies that there is no strong indication 
that the WEP will be violated. However, there is an issue that 
should be approached with care: to which accuracy can a small 
particle be considered a test particle in these theories? Could 
it be that the coupling affects our ability to treat small 
particles as test particles and neglect their contribution in 
the gravitational field? This issue, which has received 
attention within the context of GR \cite{Geroch} needs further 
investigation. 

On the other hand, there is not much to say at this point about  
the SEP. All theories with more degrees of freedom that GR 
violate the SEP, and the theory considered here is no exception, 
even without the $R$-matter coupling. Obviously, constraints can 
be imposed by considering the evolution of the effective 
gravitational coupling. The $R$-matter coupling should affect 
gravitational experiments, such as Cavendish experiments. An interesting point is that, since the coupling depends on the curvature, extended bodies whose size is comparable with the radius of curvature can get seriously affected. Considering the equation of motion of extended bodies in such theories will probably provide important constraints.

Finally, one can consider the more accurately tested EEP. It 
would be interesting to see how the $R$-matter coupling affects 
local non-gravitational physics. It should be stressed that this 
coupling can not be eliminated by a field redefinition, as 
mentioned above; therefore, going to a local frame can hold 
surprises. However, it is interesting to note that if $f_2(R)$ 
is such that $f_2(0)=0$ then, assuming that spacetime is locally 
flat in the background by choosing suitable coordinates and 
treating matter as a perturbation, the effect of the coupling 
should make its presence felt only at second order and the 
relevant  term should still be suppressed by $\lambda$.

All of the above are perspectives for future work. Let us close 
by discussing how the results derived here relate to that: We 
have found that massless particles follow null geodesics as 
usual. This essentially implies that experiments using light, 
such as redshift experiments, can not constrain the theory. We 
have also shown that perfect fluid volume elements follow 
geodesics of the metric $g_{\mu\nu}$. This includes dust, which 
is usually used as an approximation to study the motion of test 
particles. This result clashes with the claim that test 
particles will be affected by the $R$-coupling, but the issue of 
how close a small particle is to 
a test particle remains open. Once more, it should be stressed  
that geodesic motion alone can not be used to draw conclusions 
about EP violations. However, one could say that our results 
hint towards the fact that violations of the EP might be more 
difficult to detect than expected.

\subsubsection{Is there an extra force that can account for dark 
matter?}

In~\cite{BBHL} the theory under investigation was put forth as a 
resolution of the dark matter problem in galaxies. It was argued 
there that the extra force could account for the flat rotational 
curves of galaxies due to a ``MONDian" behaviour. Even though in 
\cite{BBHL} the extra force was calculated for a perfect fluid, 
its exact expression was not sufficiently detailed to support 
the argument of  this MOND-like behaviour, which  
remained rather qualitative. However, our findings cast 
doubts on whether the  theory will really exhibit such a  
behaviour. As a matter of fact, if we 
model the galaxy as a perfect fluid (as the authors of 
\cite{BBHL} do), then the extra force felt by the stars 
(the fluid-elements) vanishes and no dark matter phenomenology 
occurs. Another  approximation (and a rough one as well) 
consists of considering  a star in the outskirts of the galaxy 
as a test particle moving in vacuo under the influence of the 
gravitational field of the 
galaxy. In such an approximation, however, any correction to the 
motion with respect to the one predicted by Newtonian gravity 
would not be caused by  some extra force, as 
claimed in \cite{BBHL}: it would merely be due to the fact 
that 
the vacuum spacetime around the galaxy would differ from the one 
predicted by GR due to the difference in the field equations.\footnote{There have been attempts to address the dark matter 
problem in this way even in standard $f(R)$ gravity without the 
$R$-matter coupling (see for example~\cite{Capozziello:2004us}). 
Even though in vacuo the $R$-matter coupling 
vanishes, the vacuum solutions of the theory considered here and 
those of $f(R)$ gravity may differ due to the matching with an 
interior solution.} Therefore, it seems unlikely that an extra 
force due to the $R$-matter coupling can actually explain the 
flat rotation curves of galaxies.

\section{Conclusions}

We have studied metric $f(R)$ theories of gravity which include 
an additional direct coupling between the Ricci scalar and the 
matter Lagrangian. We have shown that they can not be cast into 
the form of either a usual scalar-tensor theory, {\em i.e.}~one scalar field non-minimally coupled to gravity and minimally coupled to matter, or even a multi-scalar-tensor theory with unusual couplings between the scalar and the matter, such as those studied  in \cite{Damour:1992we,Damour:1994zq}.  Therefore, they can not 
easily be dismissed as being equivalent to a theory with known 
phenomenology. On the contrary, we argued that the $R$-matter 
coupling persists (at least implicitly as a coupling between $R$ and a non-dynamical auxiliary field algebraically related to the matter) even after field redefinitions and can not be 
eliminated by the use of conformal transformations. Hence, 
the implication of its presence should be thoroughly examined.

We took a first step towards such a study by 
examining possible deviations  of the free fall trajectories 
from geodesics (of some conformal metric). Massless fields were 
considered, including a massless scalar field and the 
electromagnetic field, but in both cases motion followed null 
geodesics. The same result was obtained for null dust. 
Remarkably, also for  a perfect fluid or a massive 
scalar, the extra force vanishes and the motion remains 
geodesic as well. It is worth noting that this does not come as 
a consequence of energy conservation, but rather as a 
consequence of the fact that the energy flow is purely 
timelike and is aligned with the 
motion.

These results seem to indicate that some of the simple tests of 
the EP might not be able to discriminate between the theory 
under consideration and theories that satisfy the EP. Phrased 
otherwise, this means that this theory might have better chances 
of escaping the tight constraints related with the EP than other 
EP-violating theories. However, the same results seem to cast 
doubts on whether this theory can actually achieve the goal for 
which it was initially put forth in Ref.~\cite{BBHL}, {\em 
i.e.},  accounting for the observed galactic dynamics without 
resorting to dark matter.

Although the last argument appears to discourage  
further study, our overall conclusions do motivate future work 
on the viability of theories with $R$-matter coupling, since, 
besides the fact that they have already been used to address 
other problems \cite{Odintsov, 
Nojiri:2006ri, ShinjiRandall, DolgovKawasaki} as 
well, they also appear to be interesting toy models which can 
help us understanding issues related to the foundations of 
gravitation theory such as the EP. 
Additionally, the results presented here are, to some extent,  
preliminary: further work is needed to completely clarify the 
motion of massive and test particles in the theory, as well as 
its implications for Local Lorenz Invariance and Local Position 
Invariance. 
One of the main steps in this direction is the study of 
the post-Newtonian limit. Future work can also include 
stability issues, the Cauchy problem, as well as further 
investigations of the usefulness of the theory as a substitute 
for dark energy and dark matter in cosmology and astrophysics.

\section*{Acknowledgements}
The authors are grateful to Ted Jacobson for stimulating discussions and to Orfeu Bertolami and Francisco Lobo for their comments on an earlier version of this manuscript.
The work of T.~P.~S.~was supported by the National Science Foundation under grant PHYS-0601800. V.~F.~acknowledges support by  
the Natural Sciences and  Engineering Research Council of Canada  
(NSERC) and by a Bishop's University Research Grant. 

\section*{Appendix}

In this appendix we re-examine  a recent claim by one of us in 
Ref.~\cite{myBBHL} that the action~(\ref{1}) could be cast into 
the  form of a scalar-tensor theory\footnote{However, this 
claim was not actually used in order to obtain the  main results 
of~\cite{myBBHL}.} by starting from the 
action 
\be
\label{star}
S^\star=\int d^4x\,  \sqrt{-g} 
\left\{\frac{f_1 (\phi )}{2} +\frac{1}{2}f'_1(R-\phi) 
+[1+\lambda 
f_2(\phi)]L_m\right\} \;,
\ee
instead of~(\ref{2a}), and attempting to show that it is 
also  dynamically equivalent to~(\ref{1}). When~(\ref{star}) is 
varied with respect to $\phi$, one obtains
\be
\label{condv}
\frac{1}{2}(R-\phi)f''_1+\lambda f'_2 L_m=0.
\ee
In the $\lambda=0$ case (or if matter is absent), and 
provided that $f''_1\neq 0$, this yields $R=\phi$ which 
straightforwardly implies that for this specific choice for 
$\lambda$ (corresponding to the usual metric $f(R)$ gravity), 
the action~(\ref{star}) becomes dynamically equivalent to the 
reduced  version of~(\ref{1}), essentially the 
action~(\ref{faction}). 
However, it was claimed in \cite{myBBHL} that this equivalence 
continues to hold when $\lambda\neq 0$ (and matter is present). 
Clearly this can not be true, unless $f'_2=0$, in which case the 
$R$-matter coupling ceases to exist. Indeed, in order for 
the action~(\ref{star}) to be dynamically equivalent 
to~(\ref{1}), 
eq. (\ref{condv}) should reduce to  $R=\phi$. This can not be 
achieved unless the second term is somehow proportional to 
$R-\phi$ and the multiplying  factor            is strictly 
non-vanishing: this would require at best some  unacceptable 
constraint on the matter Lagrangian, if possible  at all. One 
can therefore conclude that the action~(\ref{star})  can not be 
dynamically equivalent to~(\ref{1}), contrary  to what 
is claimed in \cite{myBBHL}.

As a final note on field redefinitions with the use of an 
auxiliary 
scalar, let us just mention that, instead of starting from  the 
action~(\ref{2a}), one could decide to begin from 
\be \label{2b}
S''=\int d^4x\, \sqrt{-g} \left\{ \frac{f_1( \phi)}{2}  
 +\left[  1+\lambda f_2( R) \right] L_m +\Psi(R-\phi)\right\} \;.
\ee
Varying this action with respect to $\Psi$ yields $R=\phi$ and 
one recovers the action~(\ref{1}) as before. However, in this 
case, variation with respect to $\phi$ yields
\be
\label{psi2}
\Psi=\frac{1}{2} f'_1
\ee
Assuming that $f''_1\neq 0$ and replacing eq.~(\ref{psi}) back 
in eq. (\ref{2b}) yields
\be \label{3}
S'=\int d^4x\, \sqrt{-g} \Bigg\{ \frac{f_1( \phi)}{2}
+\left[  1+\lambda f_2(R) \right] L_m+\frac{1}{2}f'_1\left( R-\phi \right)\Bigg\} \;.
\ee
This, unlike the action~(\ref{star}),  is dynamically equivalent 
to the 
action~(\ref{1}). Using the field redefinition $\Phi=f'_1(\phi) 
$ and  introducing the potential $V(\Phi)=(\phi f'_1-f_1)/2$, 
the  action~(\ref{3}) can take the form
\be
\label{final2}
S'=\int  d^4x\, \sqrt{-g} \Big\{ \frac{\Phi R}{2} -V(\Phi) + 
[1+\lambda f_2(R)] L_m \Big\} \;,
\ee
which is  a Brans-Dicke theory with $\omega_0=0$ (just like 
ordinary metric $f(R)$ gravity) but with the addition of the 
unusual $R$-matter coupling. This way one can avoid having a 
non-linear gravitational Lagrangian at the price of introducing  
 a scalar field without touching the $R$-matter coupling in any 
way.

\end{document}